\documentstyle[11pt]{article}
\begin{document}
\baselineskip= 14pt plus 1pt minus 1pt
\rightline{hep-th/9701059}
\vskip 0.7 true cm 
{\Large
\centerline{Hidden Symmetries of the Principal Chiral Model} 
\centerline{and a Nonstandard Loop Algebra}
\vskip 0.8 true cm
\centerline{C. Devchand $^1$\ \ and\ \ Jeremy Schiff $^2$}}
{\small
\vskip 0.5 true cm
\centerline{$^1$ Institut f\"ur Theoretische Physik,
Universit\"at Hannover, D-30167 Hannover, Germany }
\centerline{$^2$ Department of Mathematics and Computer Science,
Bar--Ilan University, Ramat Gan 52900, Israel}
\centerline{E-mail:\ schiff@math.biu.ac.il}}
\vskip 0.8 true cm

\noindent {\bf Abstract.\ } 
We examine the precise structure of the loop algebra of `dressing'
symmetries of the Principal Chiral Model, and discuss a new infinite set of
abelian symmetries of the field equations which preserve a symplectic
form on the space of solutions.
\vskip 0.5 true cm

{\bf 1.\ } 
The symmetries of the classical two-dimensional Principal Chiral Model,
one of the original toy models for nonabelian gauge theories, have been 
intensively studied for over 15 years. Since the symmetry algebras
are infinite dimensional, requiring their preservation at the quantum
level provides strong constraints on the quantum theory.
In particular, it has been suspected that an investigation of their
representation theory would yield some clue about the bound
state spectrum of nonabelian gauge theories.
The aims of this paper are (a) to explain the rather subtle structure of
the celebrated loop algebra of hidden symmetries of PCM, summarising
the important points of our longer paper \cite{ds}, and (b) to discuss
some new abelian symmetries of the model. We will show that the latter 
symmetries preserve a certain natural symplectic form on the space of
solutions, which leads us to expect them to be of central importance for
an algebraic quantization of the theory. Recently there has been a renewed 
effort to understand the symmetries of PCM because of their suggested 
relevance for string theory \cite{s}.

{\bf 2.\ }  
The classical two-dimensional PCM is the differential equation
\begin{equation} 
\partial_-(g^{-1} \partial_+ g) + \partial_+(g^{-1} \partial_- g) = 0\ ,
\label{pcmg}\end{equation}
where $g$ is a mapping from two-dimensional Minkowski space to U(N).
The algebraic structure of the hidden symmetry transformations 
associated with  the infinite set of non-local conserved currents 
\cite{p,lp} of the PCM was first discussed by Dolan \cite{d}, 
who determined the algebra
\begin{equation}
[J_r^a,J_s^b] = \sum_c f^{ab}_c J_{r+s}^c\  ,\qquad  r,s\ge 0\ ,
\label{kkm}\end{equation}
where $f^{ab}_c$ are structure constants of the Lie algebra
of U(N) in a basis $\{T^a\}$; 
$[T^a,T^b] = \sum_c f^{ab}_c T^c$.
By using, for the infinite set, a compact generating function form of 
transformation,
\begin{equation}
g \mapsto g \left(I-Y(x,\lambda) T Y^{-1}(x,\lambda)\right)\ ,
\label{gf}\end{equation}
Dolan's inductive arguments were streamlined in \cite{df}, 
allowing a direct verification of the closure of the commutation relations 
(\ref{kkm}) on the fields $g$. 
Here $T$ is a constant infinitesimal
antihermitian matrix and $Y(x,\lambda)$ satisfies 
the PCM Lax-pair \cite{p,zm},
\begin{equation}
\left( \partial_\pm + {1\over 1\pm\lambda}A_\pm\right) Y =0\ ,
\quad\quad\quad A_\pm = g^{-1}\partial_\pm g\ ,
\label{lax}\end{equation}
in virtue of which (\ref{gf}) may easily be shown to generate symmetries
of equation (\ref{pcmg}). $Y$ is singular on the Riemann 
sphere only at $\lambda=\pm 1$ and satisfies the 
reality/boundary conditions
\begin{eqnarray}
Y^{\dag}(\lambda^*) = Y^{-1}(\lambda)\  ,\quad
Y(x, \lambda=\infty) = I\  ,\quad
Y(x, \lambda=0) = g^{-1}\  ,\quad Y(x_0,\lambda) = I\  ,
\end{eqnarray}
where $x_0$ is some fixed point.

A contour integral representation of the transformation (\ref{gf}) is
obtained in \cite{ds}, corresponding to similar representations in the
literature (see, for example, \cite{s,un}):
\begin{equation}
g \mapsto g\left( I - \frac1{2\pi i}\int_{{\cal C}}
\frac{Y(x,\lambda')\epsilon(\lambda')Y^{-1}(x,\lambda')}{\lambda'} 
d\lambda'\right)\  ,
\label{grd}\end{equation}
where the contour ${\cal C}$ is the union of two contours ${\cal C}_\pm$
around $\lambda=\pm 1$ (such that $\lambda=0$ remains outside both of
them). If the Lie-algebra-valued
infinitesimal parameter of the transformation
$\epsilon(\lambda)$ is taken to be proportional to $\lambda^r T^a$,
$r\in{\bf Z}$, the integral may be evaluated (for $r<0$ by
deforming ${\cal C}$ to a contour around $0$; for $r>0$ to a contour
around $\infty$; and for $r=0$ to a pair of contours around $0$ and
$\infty$) and an algebra of symmetries 
\begin{equation} 
[J_r^a,J_s^b] = \sum_c f^{ab}_c J_{r+s}^c\ , \qquad r,s\in Z\ ,
\label{ukm}\end{equation} 
may be identified, verifying the determination of these commutation
relations by \cite{w}. A careful consideration of 
the symmetries (\ref{grd}) shows, however, that {\em finding the 
commutation relations (\ref{ukm}) is not sufficient to identify the 
symmetry algebra with the standard loop algebra.} In particular:

\noindent a) Whereas for the standard loop algebra we require that the 
infinite sum $\sum\alpha_{ra} J_r^a$ should be admitted if and only if 
$\sum\alpha_{ra} \lambda^rT^a$ is convergent for $\vert\lambda\vert=1$, 
(thus defining a map from the unit circle into the Lie algebra of U(N)),
the infinite linear combinations of the $J_r^a$ allowed in our case are
entirely different.

\noindent b) There are transformations of the form (\ref{grd}) which cannot
be expressed as linear combinations of the $J_r^a$.

\noindent c) Possibly most importantly, in the algebra associated with
the transformations (\ref{grd}), 
the elements $J_r^a$ with $r<0$ are in fact in
the closure of the linear span of the transformations $\{J_r^a; r\ge 0\}$.
This means that the elements with   $J_r^a$ with $r<0$ 
are not strictly linearly independent of those with $r\ge 0$.

{\bf 3.\ }
Let $G_-$ (resp. $G_+$) be the group of smooth maps from
${\cal C}$ to U(N) which are the boundaries of maps analytic
inside (resp. outside) ${\cal C}$ , i.e. analytic in the region
$\{|\lambda-1|<\delta\}\cup \{|\lambda+1|<\delta\}$ (resp. 
$\{|\lambda-1|>\delta\}\cap \{|\lambda+1|>\delta\}$), where $\delta<1$ 
is some radius. We denote the corresponding Lie algebra ${\cal G}_-$
(resp.  ${\cal G}_+$). $Y$ satisfying (\ref{lax}) clearly takes values in
$G_+\ $. 

The symmetry algebra associated with nontrivial transformations of the form
(\ref{grd}) is ${\cal G}_-\ $, a nonstandard loop algebra. This is explained
fully in \cite{ds}, but essentially it is because we clearly can take 
$\epsilon({\lambda})$ in (\ref{grd}) 
to be an arbitrary infinitesimal element of  ${\cal G}_-\ $;
taking it to be an infinitesimal element of ${\cal G}_+\ $
gives a trivial transformation.

Choosing $\epsilon({\lambda}) \in {\cal G}_-\ $,
the natural way to expand it is in a Taylor series in $\lambda+1$ 
(or alternatively in $\lambda-1$). Taking $\epsilon(\lambda)$ to be 
proportional to $(\lambda+1)^n T^a$, for $n\ge 0$, we define a set of 
transformations $\{K_n^a\}$ satisfying the algebra
\begin{equation}
[K_n^a,K_m^b] = \sum_c f^{ab}_c K_{n+m}^c\ ,\qquad  n,m\ge 0\ .
\label{Kkm}\end{equation}
The transformations $J_r^a$ satisfying the 
relations (\ref{ukm}) are obtained in terms of the $K_n^a$ 
by considering the expansion of $\lambda^r$ in powers of $\lambda+1$ 
(valid in $\vert\lambda+1\vert<\delta$). This gives 
\setcounter{equation}{0}\renewcommand\theequation{9\alph{equation}}
\begin{equation}
 J_r^a =    \sum_{n=0}^r (-1)^{n+r}
                               \pmatrix{r\cr n\cr} K_n^a\ 
                     ,\qquad\qquad   r\ge 0\  ,\label{JfKa}\end{equation}
\begin{equation}
               J_r^a  =  \sum_{n=0}^{\infty} (-1)^r
                    \pmatrix{n-r-1\cr -r-1\cr} K_n^a\ 
             ,\quad\quad   r< 0\  .\label{JfKb} 
\end{equation}
\setcounter{equation}{9}\renewcommand\theequation{\arabic{equation}}
Standard formulae for sums of binomial coefficients  
may be used to verify that the commutation relations 
(\ref{Kkm}) $\Rightarrow$ (\ref{ukm}).
Just as the $J$'s for non-negative $r$ can be expressed as finite sums
of the $K$'s,  the latter can likewise be expanded as a finite linear
combination of the former:
\begin{equation} K_n^a=\sum_{r=0}^n \pmatrix{n\cr r\cr}J_r^a\  .
\label{KfJ}\end{equation}
When we substitute (\ref{KfJ}) into the right hand side of (\ref{JfKb}),
we find that we cannot reorder the summations to express this infinite 
sum as a linear combination of the  $J_r^a$'s with $r\ge 0$. 
In other words, if in the standard loop algebra we define elements
$K_n^a$ via (\ref{KfJ}), the infinite sum on the RHS of (\ref{JfKb})
is not in the algebra, while it is in the nonstandard loop
algebra ${\cal G}_-\ $.

In general, infinite linear combinations of the $K$'s cannot be written 
as linear combinations of the  $J_r^a$. Elements of ${\cal G}_-$ can however
be {\em approximated} (to arbitrary accuracy) by finite sums of the $J_r^a$, 
$r\ge 0$, as required by  Runge's theorem (see, 
for example, \cite{r}). This notwithstanding,  the elements $J_r^a$ 
are not a spanning set for the algebra ${\cal G}_-\ $, as they are
for the standard loop  algebra; the spanning set for the algebra 
${\cal G}_-$ is the set $\{K_n^a\}$. To see 
immediately that the elements $\{ J_r^a \}$
are not a spanning set for the algebra ${\cal G}_-\ $, 
one need only consider an element of ${\cal G}_-\ $
proportional to $\ \ln\lambda$, defined with a cut from
$0$ to $\infty$ along half of the imaginary axis.

{\bf 4.} We now describe some new PCM symmetries. The PCM potentials
$A_{\pm}$ satisfy the equations of motion
\begin{equation}
\partial_{\mp}A_{\pm}=\pm{\textstyle{\frac12}}[A_+,A_-]\ . 
\label{pcmA}\end{equation}
These imply that the eigenvalues of $A_+$ (resp. $A_-$) are independent of 
$x^-$ (resp. $x^+$). In other words, $A_+$ and $A_-$ are similarity
transformations of diagonal antihermitean matrices $A(x^+)$ and $B(x^-)$
respectively,
\begin{equation}\begin{array}{rll}
A_+ &=& s_0(x^+,x^-)A(x^+)s_0^{-1}(x^+,x^-)\  ,\\[6pt]
A_- &=& \tilde{s}_0(x^+,x^-)B(x^-)\tilde{s}_0^{-1}(x^+,x^-)\ , 
\label{s2A}\end{array}\end{equation}
where $s_0,\tilde{s}_0$ are U(N)-valued fields. The construction of 
\cite{ds} produced solutions of this form, with $A,B$ free fields and 
$s_0, \tilde{s}_0$ satisfying certain equations; the equations for $s_0$ are
\begin{equation}\begin{array}{rll}
\partial_+ s_0 s_0^{-1} &=& [t,A_+]\  ,\\[6pt]
\partial_- s_0 s_0^{-1} &=& {\textstyle{\frac12}}(s_0Bs_0^ {-1}-A_-)\  ,
  \label{sev}\end{array}\end{equation}
where $t$, an auxiliary Lie-algebra-valued field, satisfies
\begin{equation}
\partial_-t = {\textstyle{\frac14}}(s_0Bs_0^ {-1}-A_-)
    +{\textstyle{\frac12}}[t,A_-]\  .\label{t}
\end{equation}
The latter equation is sufficient for the consistency of the system 
(\ref{sev}).

Now, {\it if $f(x^+)$ is an arbitrary infinitesimal 
diagonal antihermitean matrix depending only on $x^+$, the transformation
\begin{equation}
g \mapsto g\left( I + s_0 f(x^+) s_0^{-1} \right)
\end{equation}
is a symmetry of the PCM}. To prove this we note that under an arbitrary 
infinitesimal transformation of the form $g\mapsto g(I+\Phi)$, where 
$\Phi$ takes values in the Lie-algebra, we have 
$A_{\pm} \mapsto A_{\pm} + D_{\pm} \Phi$, where $D_{\pm}$ denotes the 
covaraint derivative defined by 
$D_{\pm}\Phi = \partial_{\pm}\Phi + [A_{\pm},\Phi]$.
It is straightforward to check that $\Phi=s_0f(x^+)s_0^{-1}$ satisfies
$\partial_-D_+ \Phi + \partial_+ D_- \Phi =  2\partial_- \partial_+ \Phi 
+ [A_+,\partial_- \Phi] + [A_-,\partial_+ \Phi] = 0$ 
in virtue of (\ref{pcmA},\ref{sev},\ref{t}) and therefore generates 
a symmetry. A cumbersome calculation shows that 
{\it these new symmetries form an infinite-dimensional abelian algebra}; 
this calculation is made redundant by the considerations of \cite{ds}.  

{\bf 5.} When considering symmetries of classical equations of a field 
theory, special importance is attached to symmetries (vector fields 
on the space of solutions) which preserve a symplectic form on the 
space of solutions. As noted above, associated with any solution of 
the PCM is a gauge potential with components $A_{\pm}$. On the space of 
gauge potentials on a two-manifold there is a natural symplectic
structure \cite{AB},
\begin{equation}
\omega=\int Tr\left(\delta A\wedge \delta A\right)
      =\int dx^+dx^-~Tr\left(\delta A_+ \wedge \delta A_-\right).
\label{omega}\end{equation}
This symplectic form plays a central role in Chern-Simons theory \cite{W};
see for example \cite{BN}. We consider the pullback of this symplectic
form to the space of solutions of the PCM. This has also been considered 
in \cite{ks}.

The symplectic form $\omega$ is known to be gauge invariant. Moreover, 
the potentials satisfying (\ref{pcmA}) are pure-gauge; so, in particular, 
all symmetries of the equations are gauge transformations. This seems to mean
that all symmetry transformations must leave the symplectic form invariant, 
implying that it is totally degenerate (i.e. zero). This is an incorrect
argument, however. PCM solutions form a {\it subset} of all pure gauge 
potentials and gauge transformations which are genuine PCM symmetries are
{\it field-dependent}, whereas the invariance of $\omega$ is under 
field-independent gauge transformations.

Let us consider the symplectic form $\omega$ on a space on which the $x^+$ 
coordinate is compactified, and $x^-\in [a,b]$. An infinitesimal 
transformation of PCM solutions given by $g\mapsto g(I+\Phi)$ corresponds 
to the vector field 
\begin{equation}
V=(D_+\Phi) \frac{\delta}{\delta A_+} +
  (D_-\Phi) \frac{\delta}{\delta A_-}  
\end{equation}
on the space of potentials, and we have 
\begin{equation}
i_V\omega = 
\int dx^+dx^-~Tr\left((D_+\Phi) \delta A_- - 
                      (D_-\Phi) \delta A_+ \right).
\end{equation}
Integrating by parts, and using the fact that all PCM potentials have
zero curvature, which means that $D_+\delta A_- - D_-\delta A_+=0$, we obtain
\begin{equation}
i_V\omega = 
\left. \int dx^+~Tr(\Phi \delta A_+) \right\vert_{x^-=b}^{x^-=a}~~.
\end{equation}
Using the first equation of (\ref{s2A}) to write $A_+$ in terms of
$s_0$ and $A$, this yields
\begin{equation}
i_V\omega = 
\left. \int dx^+~Tr\left(s_0^{-1}\Phi s_0 \delta A
   + [A,s_0^{-1}\Phi s_0] s_0^{-1} \delta s_0 
   \right) \right\vert_{x^-=b}^{x^-=a}~~.
\end{equation}
For the new PCM symmetries described in section 4, we have 
$\Phi = s_0f(x^+)s_0^{-1}$, where $[f,A]=0$, and $f$ is field
independent. It follows that for such  
symmetries we have $i_V\omega=\delta \int dx^+ Tr\left(f(x^+)A(x^+)\right)$,
implying that $\delta (i_V\omega) = 0$, i.e. the symplectic form is 
preserved under these symmetries. 

The new PCM symmetries of section 4 thus preserve a symplectic form. The 
loop algebra symmetries are not believed to have such a property (see, 
e.g., \cite{Davies}). Our symplectic structure on the space of PCM 
solutions is not the standard one. Usually the symplectic form is 
derived from a Lagrangian, and  the standard  PCM Lagrangian does not
give the above symplectic form. If, however, in the Lagragian approach, 
one of the light-cone coordinates is regarded as `time', the 
symplectic forms coincide. (This refers to the standard Lagrangian 
for the PCM, not the so-called `dual' formulation \cite{zm}.) 
Although this choice appears not to be a `physical' one,
we have some hopes 
that the new symmetries described in this letter, as well as
the other constructions of \cite{ds}, will shed some light on algebraic
quantization of the PCM.

\noindent{\it Acknowledgments.}
Most of the work reported here was performed while one of us (CD) 
visited Bar--Ilan University. He thanks the Emmy Noether Mathematics 
Institute there for generous hospitality.

\goodbreak
\end{document}